\documentclass[aps,prl,twocolumn,showpacs,superscriptaddress,showkeys,bm,amsmath,amssymb]{revtex4-1}
\usepackage{graphicx} 
\usepackage{xcolor}

\begin{document}

\title{Search for sterile neutrino oscillation using RENO and NEOS data}
\newcommand{\CNU}{\affiliation{Institute for Universe and Elementary Particles, Chonnam National University, Gwangju 61186, Korea}}
\newcommand{\DSU}{\affiliation{Institute for High Energy Physics, Dongshin University, Naju 58245, Korea}}
\newcommand{\GIST}{\affiliation{GIST College, Gwangju Institute of Science and Technology, Gwangju 61005, Korea}}
\newcommand{\IBS}{\affiliation{Institute for Basic Science (IBS), Daejeon 34126, Korea}}
\newcommand{\KAIST}{\affiliation{Department of Physics, Korea Advanced Institute of Science and Technology, Daejeon 34141, Korea}}
\newcommand{\KNU}{\affiliation{Department of Physics, Kyungpook National University, Daegu 41566, Korea}}
\newcommand{\SNU}{\affiliation{Department of Physics and Astronomy, Seoul National University, Seoul 08826, Korea}}
\newcommand{\SYU}{\affiliation{Department of Fire Safety, Seoyeong University, Gwangju 61268, Korea}}
\newcommand{\SKKU}{\affiliation{Department of Physics, Sungkyunkwan University, Suwon 16419, Korea}}
\newcommand{\SJU}{\affiliation{Department of Physics and Astronomy, Sejong University, Seoul 05006, Korea}}
\newcommand{\KAERI}{\affiliation{HANARO Utilization Division, Korea Atomic Energy Research Institute, Daejeon 34057, Korea}}
\newcommand{\CAU}{\affiliation{Department of Physics, Chung-Ang University, Seoul 06974, Korea}}
\newcommand{\CUPIBS}{\affiliation{Center for Underground Physics, Institute for Basic Science (IBS), Daejeon 34126, Korea}}
\newcommand{\UST}{\affiliation{IBS School, University of Science and Technology (UST), Daejeon, 34113, Korea}}
\newcommand{\KRISS}{\affiliation{Korea Research Institute of Standards and Science, Daejeon 34113, Korea}}
\newcommand{\KUACC}{\affiliation{Department of Accelerator Science, Korea University, Sejong 30019, Korea}}

\author{Z.~Atif}\CNU
\author{J.~H.~Choi}\DSU
\author{B.~Y.~Han}\KAERI
\author{C.~H.~Jang}\CAU
\author{H.~I.~Jang}\SYU
\author{J.~S.~Jang}\GIST
\author{E.~J.~Jeon}\CUPIBS
\author{S.~H.~Jeon}\SKKU
\author{K.~K.~Joo}\CNU
\author{K.~Ju}\KAIST
\author{D.~E.~Jung}\SKKU
\author{B.~R.~Kim}\CNU
\author{H.~J.~Kim}\KNU
\author{H.~S.~Kim}\SJU
\author{J.~G.~Kim}\SKKU
\author{J.~H.~Kim}\SKKU
\author{J.~Y.~Kim}\SJU
\author{J.~Y.~Kim}\CNU
\author{S.~B.~Kim}\SKKU
\author{S.~Y.~Kim}\SNU
\author{W.~Kim}\KNU
\author{Y.~D.~Kim}\CUPIBS\UST
\author{Y.~J.~Ko}\CUPIBS
\author{E.~Kwon}\SKKU
\author{D.~H.~Lee}\SNU
\author{H.~G.~Lee}\SNU
\author{J.~Lee}\CUPIBS
\author{J.~Y.~Lee}\KNU
\author{M.~H.~Lee}\CUPIBS\UST
\author{I.~T.~Lim}\CNU
\author{D.~H.~Moon}\CNU
\author{Y.~M.~Oh}\CUPIBS
\author{M.~Y.~Pac}\DSU
\author{H.~K.~Park}\KUACC
\author{H.~S.~Park}\KRISS
\author{J.~S. Park}\KNU
\author{K.~S.~Park}\CUPIBS
\author{R.~G.~Park}\CNU
\author{H.~Seo}\SNU
\author{J.~W.~Seo}\SKKU
\author{K.~M.~Seo}\SJU
\author{C.~D.~Shin}\CNU
\author{K.~Siyeon}\CAU
\author{G.~M.~Sun}\KAERI
\author{B.~S.~Yang}\SNU
\author{I.~S.~Yeo}\SYU
\author{J.~Yoo}\SNU
\author{S.~G. Yoon}\SNU
\author{I.~Yu}\SKKU
\collaboration{The RENO and NEOS Collaborations}
\date{\today}
\begin{abstract}
{We present a nearly reactor model independent search for sterile neutrino oscillation using 2\,509\,days of RENO near detector data and 180 days of NEOS data. The reactor related systematic uncertainties are significantly suppressed as both detectors are located at the same reactor complex of Hanbit Nuclear Power Plant. The search is performed by electron antineutrino\,($\overline{\nu}_e$) disappearance between six reactors and two detectors with flux-weighted baselines of 419\,m\,(RENO) and 24\,m\,(NEOS). A spectral comparison of the NEOS prompt-energy spectrum with a no-oscillation prediction from the RENO measurement can explore reactor $\overline{\nu}_e$ oscillations to sterile neutrino. Based on the comparison, we obtain a 95\% C.L. excluded region of $0.1<|\Delta m_{41}^2|<7$\,eV$^2$. We also obtain a 68\% C.L. allowed region with the best fit of $|\Delta m_{41}^2|=2.41$\,eV$^2$ and $\sin^2 2\theta_{14}$=0.08 having a p-value of 8.2\%. Comparisons of obtained reactor antineutrino spectra at reactor sources are made among RENO, NEOS, and Daya Bay to find a possible spectral variation.}
\end{abstract}
\pacs{13.15.+g, 14.60.Pq, 14.60.St, 29.40.Mc}
\keywords{sterile neutrino, reactor antineutrino, neutrino oscillation, RENO, NEOS}
\maketitle

\par Precision measurements of the smallest neutrino mixing angle $\theta_{13}$ have established the three flavor neutrino\,(3$\nu$) oscillation framework\,\cite{RENO2012theta13,Daya2012theta13,DoubleChooz2011theta13}. However, there exist several experimental anomalies that suggest the 3$\nu$-model may not be sufficient\,\cite{LSND1996,MiniBooNE2013pmq,Gallex1998,Sage1999}. The existence of additional inactive flavor neutrinos, so-called sterile neutrinos, is often introduced to explain these anomalies.

\par The RENO and Daya Bay collaborations recently made reactor model independent searches for sterile neutrino by comparing the observed prompt spectra of near and far detectors and obtained excluded regions of sub-eV sterile neutrino oscillations\,\cite{RENO_sterile,Daya_Minos_sterile}. The NEOS collaboration reported a result of sterile neutrino search\,\cite{NEOS2016prl} using the Daya Bay's $\overline{\nu}_e$ spectrum at the reactor\,\cite{Daya_absolute}. The reference $\overline{\nu}_e$ spectrum used by NEOS was obtained by including reactor related uncertainties and assuming no sterile neutrino oscillation between reactor and detector. These uncertainties and assumption are not needed if a RENO's $\overline{\nu}_e$ spectrum is used for the comparison. By utilizing full response functions of both detectors, a robust statistical analysis is possible to obtain an exclusion limit and a p-value. A spectral comparison of the NEOS\,(RENO) prompt spectrum with the $3\nu$ best-fit prediction\,\cite{PDG2020} from the RENO\,(NEOS) measurement can explore sterile neutrino oscillations by reactor $\overline{\nu}_e$ disappearance.

\par This Letter presents a nearly reactor model independent search for sterile neutrino oscillation using 2\,509 days and 180 days of data taken in the RENO near detector and the NEOS detector, respectively. Both detectors are located at the Hanbit Nuclear Power Plant, Korea. The reactor complex hosts six pressurized water reactors. The maximum thermal output of each reactor is 2.8\,GW$_{\text{th}}$. The RENO near detector is located at 294\,m from the center of the reactor array, and the flux-weighted baseline is 419\,m.  The NEOS detector is placed in the tendon gallery of the fifth reactor. The distance between the NEOS detector and the reactor core is $23.7\pm0.3$\,m. Both detectors employ gadolinium loaded liquid scintillator to observe the inverse beta decay\,(IBD) interactions of reactor $\overline{\nu}_e$. Their $\overline{\nu}_e$ target volumes are 16.5\,tons for RENO and 0.87\,tons for NEOS. Details of RENO and NEOS experiments can be found in Refs.\,\cite{RENO2018PRD,RENO2010TDR,NEOS2016prl}.

\par The combined analysis of RENO and NEOS data removes a large part of the reactor related uncertainties. Furthermore, a sterile neutrino hypothesis can be tested using the observed spectrum at each baseline, without knowing the $\overline{\nu}_e$ non-oscillation spectrum at reactor. The survival probability of reactor $\overline{\nu}_e$ in the (3+1)\,$\nu$-oscillation hypothesis is given as\,\cite{PDG2020},
\begin{eqnarray}
&P_{\overline{\nu}_e \rightarrow \overline{\nu}_e}  =
1-4\sum_{i>j}|U_{ei}|^2|U_{ej}|^2\sin^2\Big(\frac{\Delta m_{ij}^2 L}{4E}\Big) \nonumber \\
&\simeq 1- \sin^2 2\theta_{14} \sin^2 \Delta_{41} - \sin^2 2\theta_{13} \sin^2 \Delta_{31},
\label{eq:survival probability}
\end{eqnarray}

\noindent where $|U_{ei}|$ and $|U_{ej}|$ are the elements of the neutrino mixing matrix, $L$ is the baseline between reactor and detector, $E$ is $\overline{\nu}_e$ energy, $\Delta_{ij}=\,\Delta m_{ij}^2\,L/4E$, and $\Delta m^2_{ij}$ is the mass splitting between the $i$-th and $j$-th mass eigenstates. The exact expression of the survival probability is used for this analysis. A measured sterile neutrino oscillation probability determines the mixing angle of $\theta_{14}$ and the mass squared difference of $\Delta m^2_{41}$. In the region of $\Delta m^2_{41} \sim\,$1 eV$^2$, a shorter baseline makes a larger oscillation effect as the reactor $\overline{\nu}_e$ energy is a few MeV. In the region of larger $\Delta m^2_{41}$($\gtrsim$\,3\,eV$^2$), the oscillation effect is averaged out within the volumes of reactor core and detector due to a rapid oscillation and thus hard to be measured by the NEOS detector.

To find a possible modulation in energy coming from a sterile neutrino oscillation, the prompt-energy spectrum observed in the NEOS detector is compared with the 3$\nu$ best-fit prediction at NEOS, obtained from the measurement in the RENO near detector. The spectral comparison is also performed between the prompt spectrum observed in the RENO detector and the 3$\nu$ best-fit prediction obtained from the NEOS measurement. The RENO collaboration has reported an unfolded reactor $\overline{\nu}_e$ spectrum\,\cite{RENO_absflux}. The RENO expected prompt spectrum at NEOS is obtained using the NEOS detector response function applied to the 3$\nu$ best-fit prediction from the extracted $\overline{\nu}_e$ spectrum of RENO. On the other hand, a $\overline{\nu}_e$ spectrum is also extracted by unfolding the NEOS observed prompt spectrum and is reported here for the first time. The NEOS expected prompt spectrum at RENO is obtained by applying the RENO detector response function\,\cite{RENO_absflux} to the 3$\nu$ best-fit prediction from the extracted $\overline{\nu}_e$ spectrum of NEOS.

\begin{figure}[t!]
\includegraphics[width=0.48\textwidth]{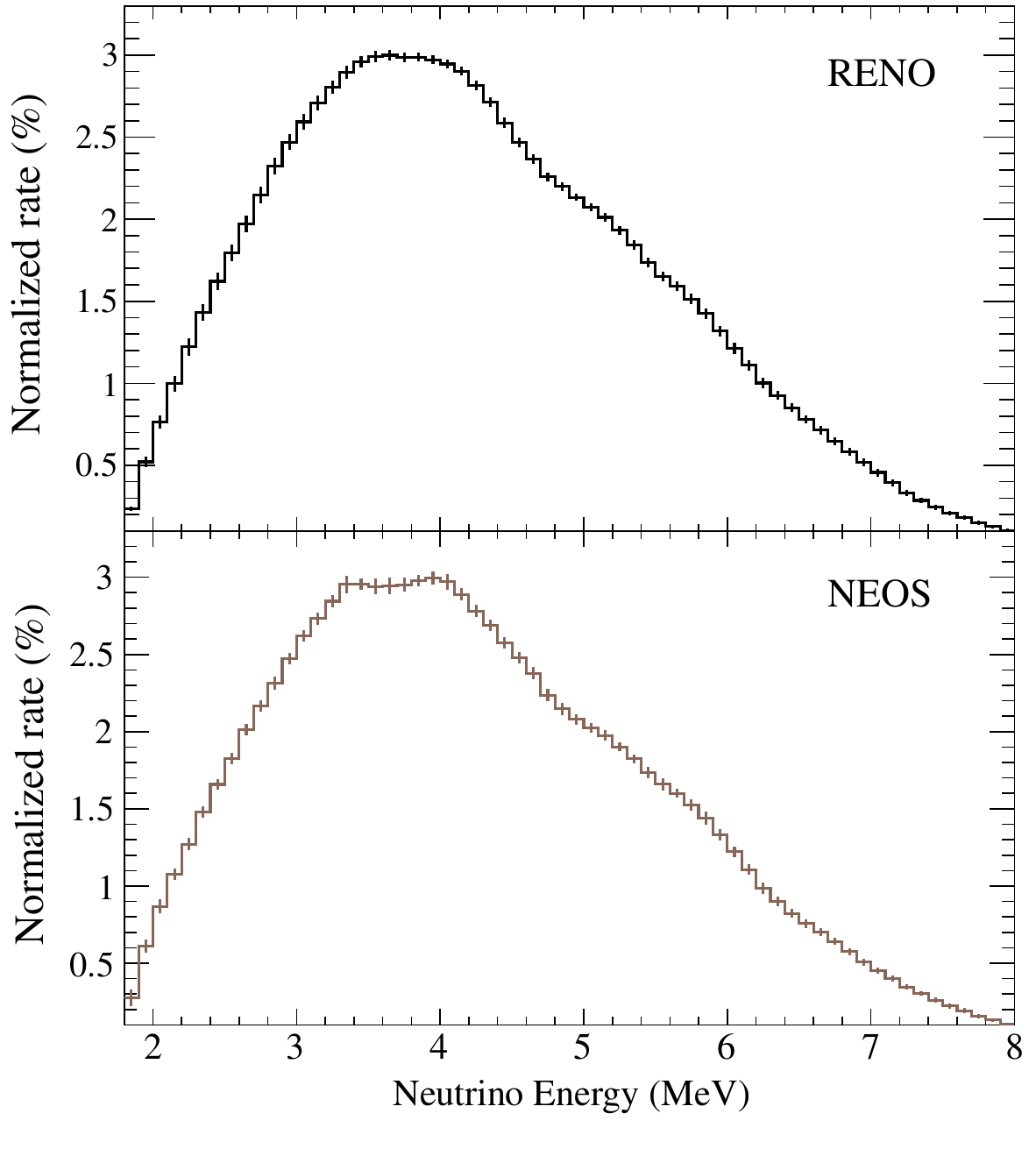} 
\caption{Reactor antineutrino spectra obtained from the measured IBD prompt spectra of RENO and NEOS after unfolding their detector effects. The error bars are obtained as square root of the diagonal elements of the corresponding covariance matrices. Each spectrum is normalized by the total integral of the spectrum.}
\label{fig:compare_deconvolution}
\end{figure}

\par The detector response effects in the prompt spectra are removed by an unfolding method of Iterative Bayesian Unfolding\,(IBU)\,\cite{IBU} which updates an initial prior spectrum by a feedback process. A number of iterations regulate weighting for an initial prior spectrum and an observed spectrum. The weighting balance is determined by the L-curve criterion\,\cite{Lcurve1995}. The fourth\,(fifth) iteration finds the best solution for the RENO\,(NEOS) data. A bias in the unfolded spectrum is introduced by uncertainties of the prompt spectrum and the unfolding algorithm. The unfolding accuracy is degraded by imperfect understanding of the detector response and a finite detector resolution. The energy-bin correlated uncertainties are included through a covariance matrix. The largest contribution to the matrix elements comes from the energy scale uncertainty in the RENO detector and the statistical fluctuation in case of NEOS. Details of the unfolding process for the RENO spectrum can be found in Ref.\,\cite{RENO_absflux}. The NEOS unfolded $\overline{\nu}_e$ spectrum is obtained in a similar way.
\begin{table}[b!]
\caption{Fractional uncertainties in the RENO and NEOS $\overline{\nu}_e$ spectra between 1.8 and 8.0\,MeV. The errors are average values in energy.}
\begin{tabular}{lcc}
\hline \hline 
		Error component \quad & \quad RENO\,(\%) \quad & \quad NEOS\,(\%) \quad \\ \hline
		Energy scale     & 1.6 & 0.9  \\
		Statistical      & 0.6 & 1.4  \\
		Unfolding        & 0.7 & 1.0  \\
		Background       & 0.3 & 0.4  \\ \hline
		Total            & 1.9 & 2.1  \\
		\hline \hline
\end{tabular}
\label{table:error}
\end{table}

\par
\begin{figure}[t!]
\includegraphics[width=0.47\textwidth]{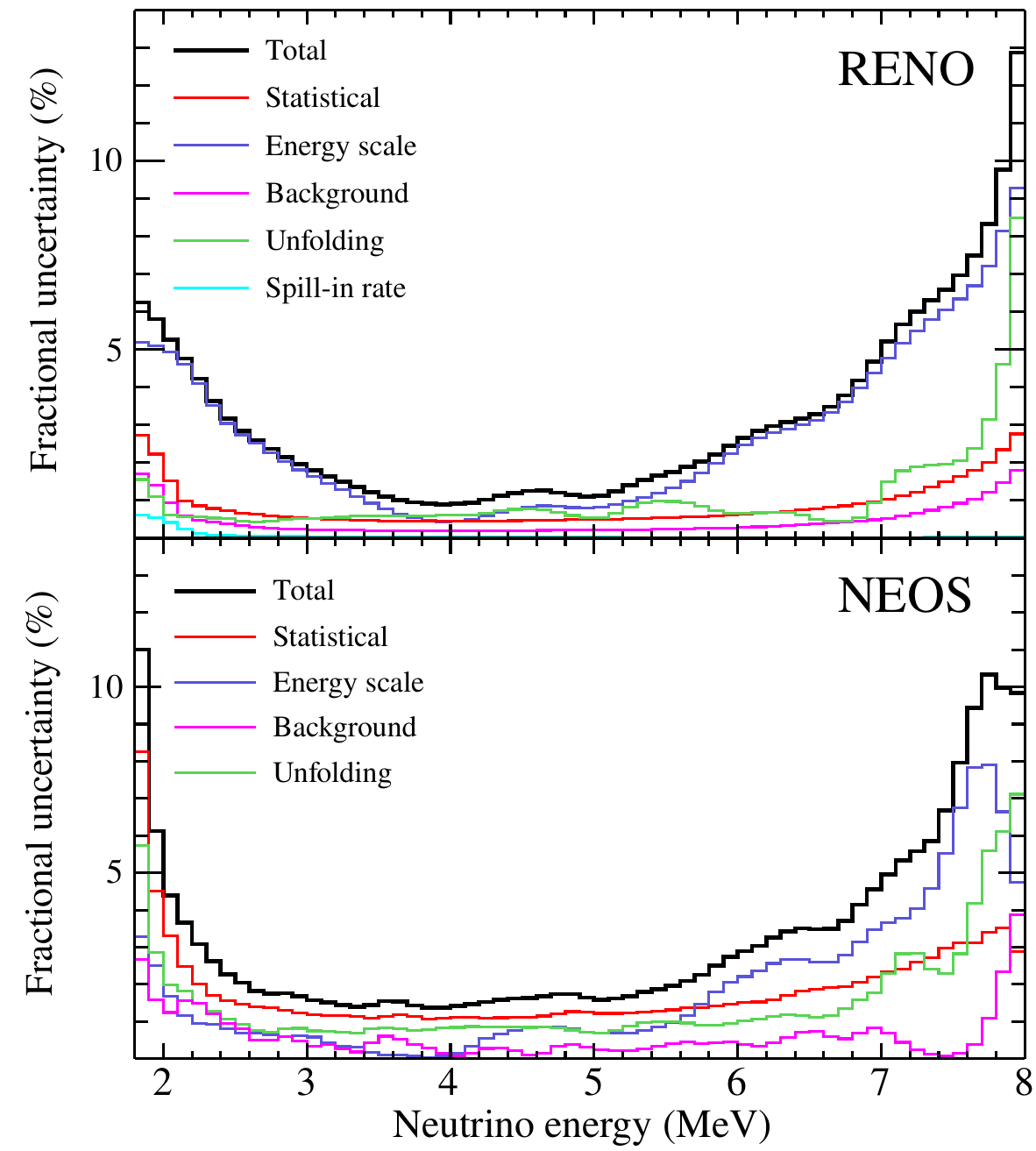}
\caption{Fractional uncertainties of unfolded RENO and NEOS $\overline{\nu}_e$ spectra.}
\label{fig:NEOS_error}
\end{figure}

 Figure\,\ref{fig:compare_deconvolution} shows the extracted RENO and NEOS $\overline{\nu}_e$ spectra with detector effects unfolded. The total uncertainty in the figure includes statistical and systematic uncertainties. However, the uncertainties of detection efficiencies are irrelevant to the shape of the spectrum and not included in the oscillation analysis. Table\,\ref{table:error} lists the fractional uncertainties of RENO and NEOS $\overline{\nu}_e$ spectra in the energy range between 1.8 and 8.0\,MeV. The various uncertainty sources are shown as a function of $\overline{\nu}_e$ energy in Fig.\,\ref{fig:NEOS_error}.

\par
\begin{figure}[t!]
\includegraphics[width=0.495\textwidth]{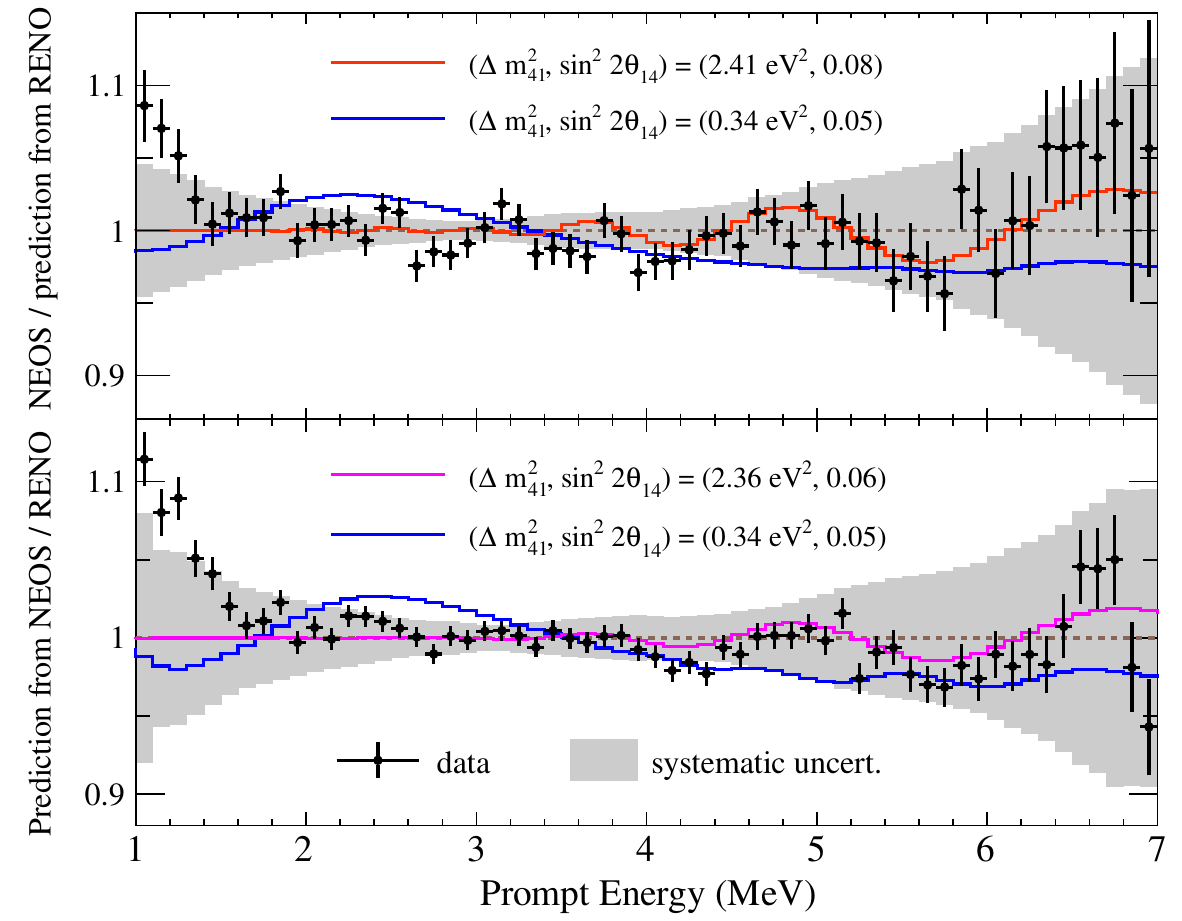}
\caption{Comparison of prompt-energy spectra at NEOS and RENO. Upper: Ratio of the NEOS observed prompt spectrum relative to the $3\nu$ best-fit prediction at NEOS from the RENO measurement. The error bars represent the statistical uncertainty only. The error bands represent the systematic and prediction uncertainties. The areas of two spectra are normalized for a shape comparison. The gray band indicates the systematic uncertainty. Lower: Ratio of the NEOS $3\nu$ best-fit prediction at RENO relative to the RENO observed prompt spectrum. The red and magenta curves represent the best fits to the data. The blue curves represent spectral ratios expected with one of sterile neutrino oscillation parameters that are excluded by this analysis.}
\label{fig:compare_ratio}
\end{figure}

 The extracted $\overline{\nu}_e$ spectra are corrected for different fuel isotope fractions between RENO and NEOS, caused by their mismatched data-taking periods and their different detector locations with respect to reactors. The correction is made using the Huber and Mueller\,(HM) predicted spectra\,\cite{Huber2011,Mueller2011}. The average fission fractions of $^{235}$U, $^{238}$U, $^{239}$Pu, and $^{241}$Pu are 0.571 (0.655), 0.073 (0.072), 0.300 (0.235), and 0.056 (0.038), respectively, for RENO (NEOS). The expected prompt spectra at RENO and NEOS are obtained from the 3$\nu$ best-fit predicted $\overline{\nu}_e$ spectra from the NEOS and RENO measurements, respectively, using their detector response functions. Figure~\ref{fig:compare_ratio} shows comparisons of prompt spectra at RENO and NEOS. The upper panel of Fig.~\ref{fig:compare_ratio} shows the NEOS observed prompt spectrum divided by the RENO prediction at NEOS, and the lower panel shows the NEOS prediction at RENO divided by the RENO observed prompt spectrum. The areas of two spectra are normalized for spectral shape comparison. The uncertainties of fuel isotope rates are correlated between RENO and NEOS. Other uncertainties are assumed to be fully uncorrelated between the RENO and NEOS spectra in the spectral comparisons.

\par A method of $\Delta \chi^2$ is chosen by this search for the sterile neutrino oscillation. A $\chi^2$ function of the spectral comparison at the NEOS detector is constructed as,
\begin{eqnarray}
\chi^2 &=\sum^N_{i,j} \Big( N_{\text{R}}^i- \alpha\frac{M^i_{\text{R}}}{M^i_{\text{N}}}N_{\text{N}}^i \Big) 
		V_{ij}^{-1}\Big( N_{\text{R}}^j- \alpha\frac{M^j_{\text{R}}}{M^j_{\text{N}}}N_{\text{N}}^j \Big),
\label{eq:chi2_detail}
\end{eqnarray}
\noindent where $N^i_{\text{R}}$ and $N^i_{\text{N}}$ are the numbers of observed events in the $i$-th $\overline{\nu}_e$ energy bin at RENO and NEOS, respectively, $M^i_{\text{R}}$ and $M^i_{\text{N}}$ are the numbers of events expected from a sterile neutrino oscillation, $\alpha$ is a scale factor for the shape comparison, and $V_{ij}$ is a covariance matrix element for a total spectral error of RENO and NEOS in the $i$-th and $j$-th $\overline{\nu}_e$ energy cell. The matrix element is given by,
\begin{eqnarray}
V_{ij} = V^{ij}_{\text{R}}+ \alpha^2 \Big( \frac{M^i_{\text{R}}}{M^i_{\text{N}}} \Big) \cdot \Big( \frac{M^j_{\text{R}}}{M^j_{\text{N}}} \Big) V^{ij}_{\text{N}},
\label{eq:covaraince_combined}
\end{eqnarray}    

\noindent where $V^{ij}_{\text{R}}$ and $V^{ij}_{\text{N}}$ are covariance matrix elements of RENO and NEOS, respectively. The value of $\chi_{3\nu}^2/$NDF for $3\nu$ oscillation parameters is 56.6/59 where NDF is the number of degrees of freedom. The minimum $\chi^2_{4\nu,\text{min}}$/NDF value for the sterile neutrino oscillation is 48.2/57. The 3$\nu$ oscillation parameters in Ref. \cite{PDG2020}, $\sin^2 \theta_{13} = (2.20 \pm 0.07) \times 10^{-2}$ and $\Delta m^2 = (2.45 \pm 0.03) \times 10^{-3} \text{eV}^2$ assuming the normal ordering, are used in the fit. The best fit shown in Fig.~\ref{fig:compare_ratio} is found at $|\Delta m^2_{41}|=2.41\,\text{eV}^2$ and $\sin^2 2\theta_{14}=0.08$. The spectral ratio of data appears to be consistent with the best-fit expectation including the energy modulation. The value of $\Delta \chi^2 = \chi^2_{3\nu} - \chi^2_{4\nu,\text{min}}$ is 8.4 and the p-value of inconsistency with the $3\nu$-model is estimated to be 8.2\,\% by generating a large number of pseudo-experiment sets. This breaks down Wilks' theorem\,\cite{Wilks:1938dza,Coloma:2020ajw,Giunti:2021iti}, which assumes that the $\Delta \chi^2$ follows the $\chi^2$ distribution. The obtained $\Delta \chi^2$ and p-value correspond to four effective degrees of freedom. Note that the $\Delta \chi^2$ and p-value obtained in Ref.\,\cite{NEOS2016prl} are 6.5 and 22\%, respectively.
\begin{figure}[t!]
\includegraphics[width=0.49\textwidth]{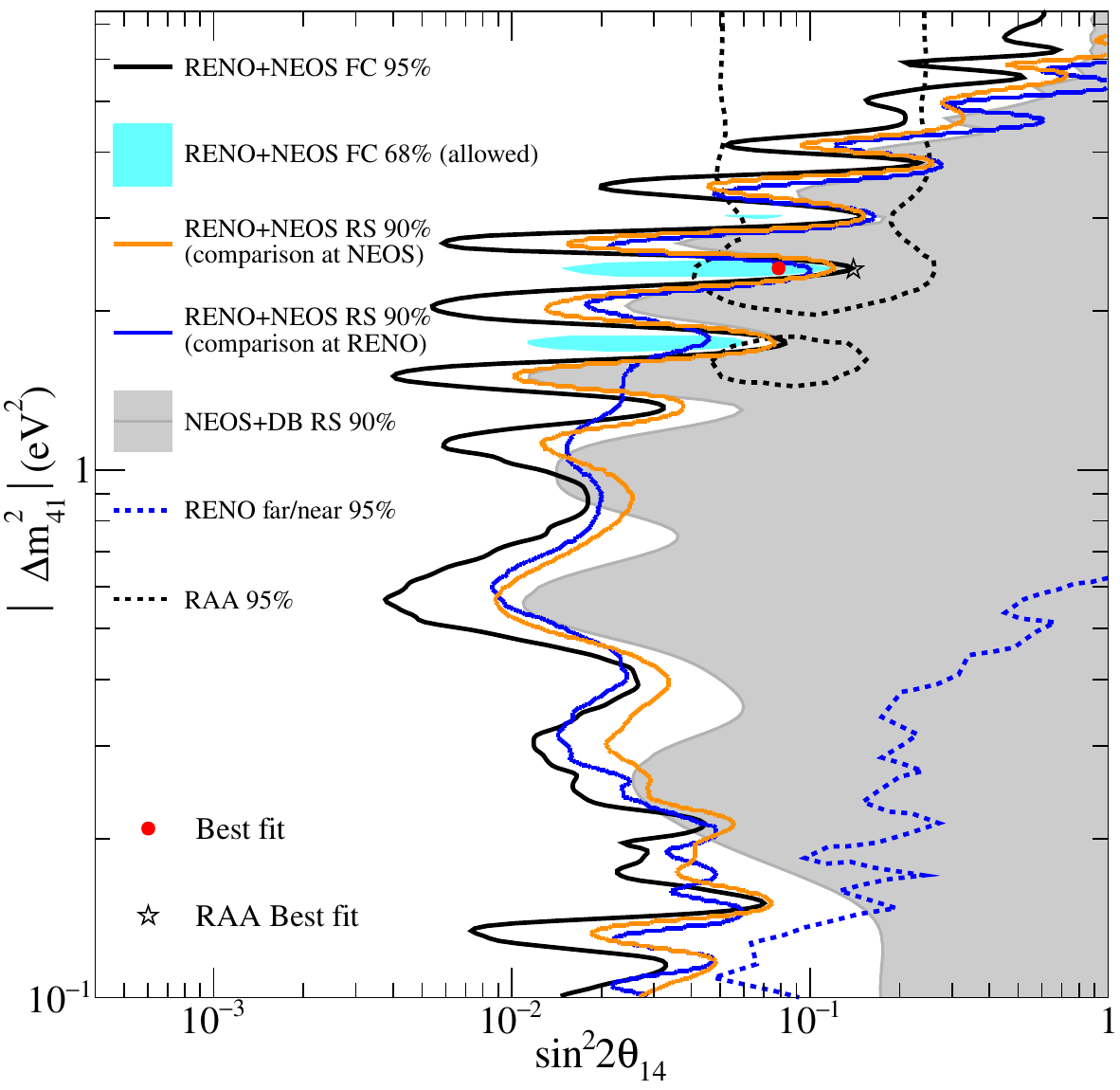}
\caption{Comparison of the exclusion limits on sterile neutrino oscillations and an allowed region. The right side of each contour indicates an excluded region. The black curve\,(cyan filled region) represents a 95\%\,(68\%) C.L. exclusion contour\,(allowed region) obtained from the RENO and NEOS combined search using the Feldman and Cousins method\,(FC)\,\cite{FeldmanCousins}. The orange\,(blue) curve represents a 90\% C.L. exclusion contour obtained from the RENO and NEOS combined search using the raster scan\,(RS) method where the spectral comparison is made at NEOS\,(RENO) detector. The best fit parameter (red point) is found at $|\Delta m^2_{41}|=2.41\,\text{eV}^2$ and $\sin^2 2\theta_{14}=0.08$. The second best-fit is found at $|\Delta m^2_{41}|=1.75\,\text{eV}^2$ and $\sin^2 2\theta_{14}=0.05$. For the comparison, shown are the NEOS+Daya Bay\,\cite{NEOS2016prl} 90\% C.L. (gray shaded) and RENO far/near\,\cite{RENO_sterile} 95\% C.L. (blue dotted) limits on the disappearance. Also shown is a 95\% C.L. allowed region of RAA\,\cite{RAA2011} (black dotted) with the best fit\,\cite{white_paper_sterile2012} (star) at $|\Delta m^2_{41}|=2.4\,\text{eV}^2$ and $\sin^2 2\theta_{14}=0.14$. Note that part of the best-fit area is excluded by STEREO and PROSPECT within 95\% C.L.\,\cite{STEREO2020,Prospect2019}.}
\label{fig:exclusion_contour}
\end{figure}

\par For a direct comparison of this result with the previous NEOS result\,\cite{NEOS2016prl}, a method of raster scan\,\cite{RasterScan2014} is used to obtain an excluded region of sterile neutrino oscillation parameters. For each set of $\sin^2 2\theta_{14}$ and $|\Delta m^2_{41}|$, $\Delta \chi^2 = \chi^2 - \chi^2_{\text{min}}$ is calculated where $\chi^2_{\text{min}}$ is the minimum $\chi^2$. In the raster scan, $\chi^2_{\text{min}}$ is determined by varying the $\sin^2 2\theta_{14}$ for an every value of $|\Delta m^2_{41}|$, unlike the 2-D scan. The parameter sets of $\sin^2 2\theta_{14}$ and $|\Delta m^2_{41}|$ are excluded at 90\% confidence level\,(C.L.) if $\Delta \chi^2$ is greater than 2.71\,\cite{RasterScan2014}. Figure~\ref{fig:exclusion_contour} shows exclusion contours obtained from the RENO and NEOS data. It also shows a 95\% C.L. allowed region of reactor antineutrino anomaly (RAA)\,\cite{RAA2011}. The excluded region is enlarged from the previous NEOS result\,\cite{NEOS2016prl}. This improvement is given by the spectral comparison based on the RENO $\overline{\nu}_e$ spectrum at the same reactor complex. The reduced systematic uncertainty using the RENO's $\overline{\nu}_e$ spectrum with a smaller uncertainty than that of Daya bay allows an additional region of $\sin^2\theta_{14}$ to be excluded in this analysis. The extended sensitivity in the $|\Delta m^2_{41}| < 0.5$\,eV$^2$ region comes from the longer baseline of the RENO near detector than NEOS. Note that the previous NEOS result\,\cite{NEOS2016prl} was based on the Daya Bay's $\overline{\nu}_e$ spectrum expected at reactor by assuming no sterile neutrino oscillation between reactor and detector\,\cite{Daya_absolute}.

\par To properly treat the confidence interval in the exclusion contour the Feldman and Cousins\,(FC) method\,\cite{FeldmanCousins} is applied in the analysis as well. The FC method rules out additional regions in the sterile neutrino parameter space with 95\,\% C.L. because of finding excluded parameters relative to the global best-fit in a two-dimensional space\,\cite{FeldmanCousins,RasterScan2014}. This improvement is equivalent to roughly 15 times NEOS statistics compared to the previous application of the raster scan method.

\par A search is repeated with the spectral comparison at RENO. The ratio between the RENO observed prompt spectrum and the NEOS prediction at RENO, as shown in the lower panel of Fig.~\ref{fig:compare_ratio}, obtains a similar exclusion contour. The best fit is found at $|\Delta m_{41}^2|=2.36$\,eV$^2$ and $\sin^2 2\theta_{14}=0.06$, consistent with those of comparison at NEOS. The value of $\Delta \chi^2$ is 4.1 where the values of the best-fit $\chi^2$/NDF are 32.7/57 and 36.9/59, respectively, for the sterile neutrino oscillation and null hypotheses. The comparison at RENO provides a more sensitive search in the sub-eV region due to its longer baseline.

\par As the spectral ratio of RENO and NEOS data appears to be consistent with the best-fit prediction within their uncertainties in Fig.~\ref{fig:compare_ratio}, allowed parameter regions at 68\,\% C.L. are obtained using the FC method, consistent with the allowed region of RAA by Ref.\,\cite{RAA2011}. The best fit is found at $|\Delta m_{41}^2|=2.41$\,eV$^2$ and $\sin^2 2\theta_{14}$=0.08, whereas that of RAA at $|\Delta m^2_{41}|=2.4\,\text{eV}^2$ and $\sin^2 2\theta_{14}=0.14$\,\cite{white_paper_sterile2012}. They are consistent with each other at $1.9\,\sigma$, obtained by a Monte Carlo simulation assuming a true signal. An improved measurement of reactor $\overline{\nu}_e$ flux and energy spectrum with a substantially reduced systematic uncertainty is essential to obtain a conclusive result on the sterile neutrino oscillation.

\begin{figure}[t!]
\includegraphics[width=0.49\textwidth]{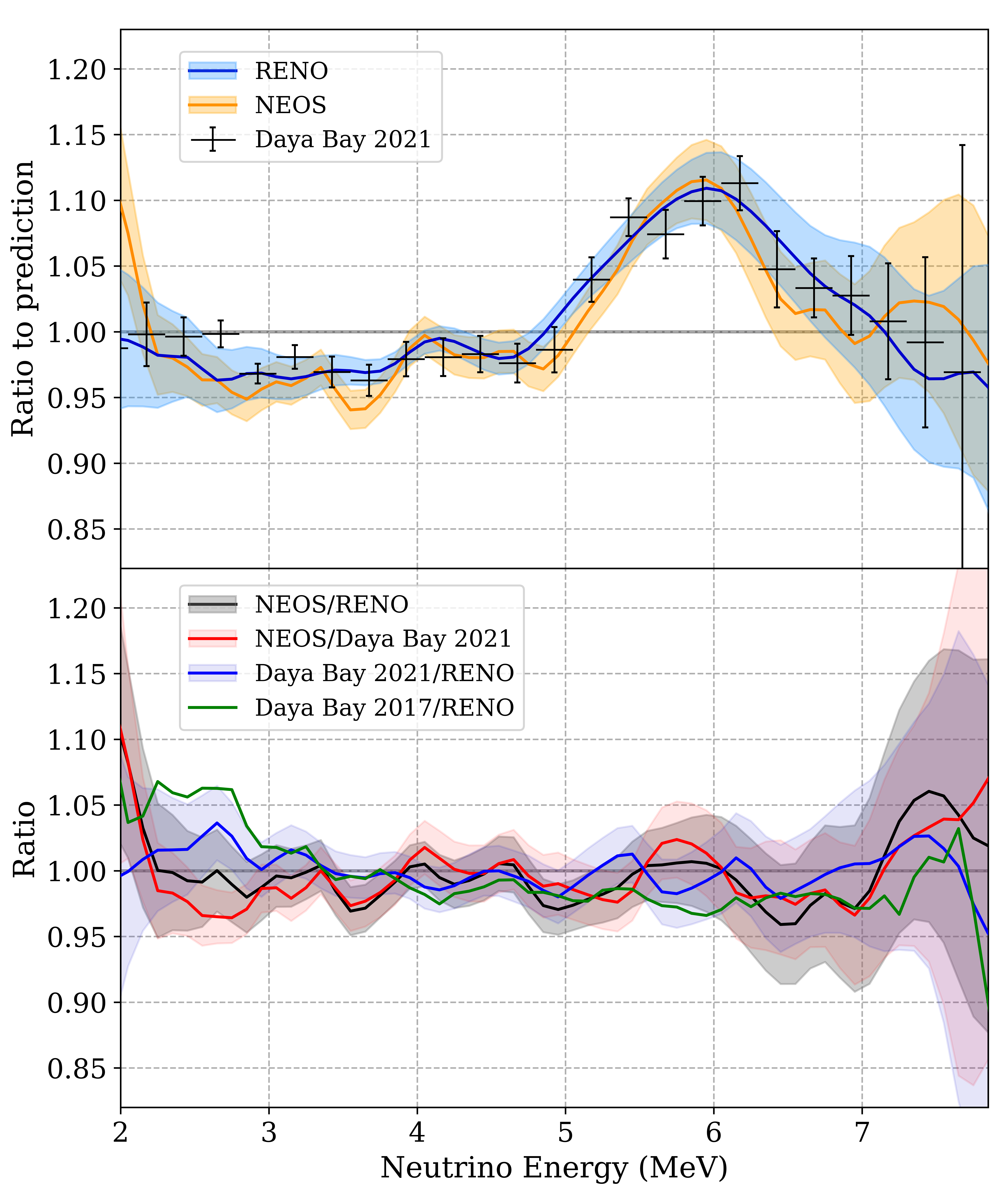}
\caption{(Upper) Comparisons of $\overline{\nu}_e$ spectral shapes of RENO, NEOS, and Daya Bay\,\cite{Adey:2021rty} relative to the HM model. The extracted $\overline{\nu}_e$ spectra are extrapolated to the production point at the reactor using the 3$\nu$ oscillation model and the parameter values in Ref.\,\cite{PDG2020}. The differences of fission fractions are corrected. Each spectrum is normalized by the area between 1.8-8\,MeV. (Lower) Comparisons of $\overline{\nu}_e$ spectral shapes with an arbitrary normalization. The Daya Bay spectrum reported in 2017\,\cite{Daya_absolute} is recently updated\,\cite{Adey:2021rty}.}
\label{fig:compare_ratio_Daya}
\end{figure}

The extracted $\overline{\nu}_e$ spectra of RENO, NEOS, and Daya Bay\,\cite{Daya_absolute,Adey:2021rty} are converted to the 3$\nu$ non-oscillating spectra at reactor using global best-fit parameters\,\cite{PDG2020}. Figure~\ref{fig:compare_ratio_Daya} shows their spectral shape comparisons with respect to the HM model and comparisons among them, exhibiting interesting similarities and discrepancies among each other. Note that Daya Bay collaboration recently updated their $\overline{\nu}_e$ spectrum\,\cite{Adey:2021rty}. The four spectra in general show agreement within their systematic uncertainties. A $\chi^2$ test of an identical spectral hypothesis between any two of these spectra gives 45.8, 36.2, 32.1, and 32.6 for NEOS/RENO, NEOS/Daya Bay 2021, Daya Bay 2021/RENO, and Daya Bay 2017/RENO, respectively, with 57 degrees of freedom. An arbitrary normalization is made with a free parameter.  A slope, probably due to their energy scale difference, is seen in the spectral comparison between Daya Bay 2017 and RENO, but Daya Bay 2021 does not show such a shift any more. The reactor related uncertainty is 0.04\% in the spectral comparison between RENO and NEOS, while it is 0.3\% between Daya Bay and one of the other two. The minimal spectral discrepancy is observed between RENO and Daya Bay. NEOS shows a weak spectral modulation analogous to the expectations from the (3+1)\,$\nu$-oscillation within the systematic uncertainty in Fig.~\ref{fig:compare_ratio}. Because of the minimal difference between RENO and Daya Bay, the apparent modulation possibly comes from the NEOS obtained $\overline{\nu}_e$ spectrum as a result of an unknown uncertainty in the measurement, a sub-structure of the reactor $\overline{\nu}_e$, or active-to-sterile neutrino oscillations. Significant improvement of systematic uncertainties and an energy resolution is required for complete understanding of the observed differences in the $\overline{\nu}_e$ spectra.

\par In summary, we report a nearly reactor model independent search for sterile neutrinos using data from the RENO near detector and the NEOS detector. As the RENO and the NEOS detectors share the same reactor complex, this analysis removes the $\overline{\nu}_e$ source dependency in the previous NEOS sterile neutrino search\,\cite{NEOS2016prl}. This analysis uses IBD spectra measured at RENO and NEOS detector locations without knowing a non-oscillation spectrum at reactor. Based on the spectral comparison of the NEOS prompt-energy spectrum with a prediction from the RENO measurement, we obtain a 95\% C.L. excluded region of $0.1<|\Delta m_{41}^2|<7$\,eV$^2$. We also obtain an allowed region with the best fit of $|\Delta m_{41}^2|=2.41$\,eV$^2$ and $\sin^2 2\theta_{14}$=0.08. Comparisons of obtained reactor antineutrino spectra at reactor sources are made among RENO, NEOS, and Daya Bay and show an interesting spectral variation within their systematic uncertainties. The reactor model independent reactor neutrino spectra in this report will be useful studying reactor neutrino physics and particle physics beyond the Standard Model.\\

\par The RENO experiment is supported by the National Research Foundation of Korea (NRF) grants No. 2009-0083526, No. 2019R1A2C3004955, and 2021R1A2C1013661 funded by the Korea Ministry of Science and ICT. The NEOS experiment is supported by IBS-R016-D1 and 2012M2B2A6029111 from National Research Foundation (NRF). Some of us have been supported by a fund from the BK21 of NRF. This work is also supported by the New Faculty Startup Fund from Seoul National University. We gratefully acknowledge the cooperation of the Hanbit Nuclear Power Site and the Korea Hydro \& Nuclear Power Co., Ltd. (KHNP). We thank KISTI for providing computing and network resources through GSDC, and all the technical and administrative people who greatly helped in making this experiment possible.  
\bibliography{renoneos_sterile}

\end{document}